# Wireless Sensing of Temperature, Strain and Crack Growth in Metal Structures via Magneto-responsive Inclusions

Connor G. McMahan[1], Chia-Ming Chang[1], Raymond Nguyen[1], Souren Soukiazian[1], David A. Smith[1], Darby LaPlant[1], Tobias Schaedler[1], David Shahan[1]*

[1]HRL Laboratories, LLC; 3011 Malibu Canyon Road, Malibu, CA 90265, USA

*Corresponding author. Email: dwshahan@hrl.com

**Abstract:** This study demonstrates the first realization of wireless strain, temperature and crack growth sensing within 3D-printed metallic structures using standard electromagnetic inspection hardware. This establishes a path toward need-based maintenance for parts operating in harsh environments driven by accurate, real-time damage assessments instead of relying on regularly scheduled maintenance teardowns. To this end, we encapsulate and embed magnetoelastic and thermomagnetic materials during additive manufacturing. Mechanical and thermal stimuli affect the magnetic permeability of the embedded materials, which modulates the flux through a coil placed on or near the part's surface. We demonstrate strain sensing accurate to $\pm 27\times10^{-6}$ and temperature sensing accurate to ±0.75 ºC. We highlight these sensors' capabilities by detecting the onset of plasticity and fatigue-driven crack growth thousands of cycles before critical failure.

## Main Text

The standard process for ensuring the safe operation of fatigue-limited structures (e.g., aircraft, automobile, and power plant components) is to perform regularly scheduled maintenance teardowns. This is costly, labor-intensive, and requires downtime that can exceed the duration of structural duty cycles. Reliable in-situ damage assessments of structural health would eliminate the need for inspecting still-intact components by providing warnings of sub-critical crack growth well in advance of catastrophic failure, but several barriers to widespread adoption of in-service structural health monitoring have not been addressed by existing technology. Most monitoring solutions today rely on wired, surface-mounted sensors that are prone to damage, difficult to install, and cannot be used in extreme environments where they could provide the most value due to tear-off and degradation. Accurate sensors that can be embedded within components for protection from harsh operating environments and that can be queried via wireless, non-contact receivers would enable a significant shift from schedule-based maintenance paradigms toward need-based intervention approaches.

Existing methods for monitoring strain, temperature and damage do not offer these capabilities. Fluorescent penetrant inspection (FPI) is the most used non-destructive method during schedule-based assessments of structural health in the aerospace industry. While significant advances have been made toward automating the optical identification of crack-permeating dye *(1–3)*, FPI often requires structural disassembly to access the inspection site. Eddy current techniques can be used to detect defects, but present challenges in signal interpretation (distinguishing cracks from material inhomogeneities), calibration (measurements are very sensitive to probe position relative to the surface), and with the geometry of the probed parts (complex features affect the measured signal) *(4, 5)*. Strain gauges and comparative vacuum monitoring (CVM) systems can be placed on the surface of a component *(6)*, but they can be expensive to install and are subject to tear-off in many operating conditions. Optical fibers that rely on Bragg gratings are compelling alternatives to conventional monitoring techniques, as they are proven mechanisms for strain measurements that can be decoupled from thermal effects with high accuracy and resolution *(7–12)*. However, probing strain fields via optical fibers requires routing them through the part to the surface (inducing relatively large defects in the part which may affect structural integrity). The fibers are also fragile and require special interfaces for sensing strain in rotating components. Similar challenges exist for temperature sensing, where the state-of-the-art technique for remote measurements is pyrometry, but this method is limited to sensing surface temperatures, while sensitive regions are often on the interior of the component. Thermocouples can be embedded within a component but require routing wires to the surface for signal read-out. As with optical fiber strain sensing, these are especially challenging to incorporate in rotating components where non-contact sensing is needed during operation.

We present a novel strain, temperature and damage sensing technology which provides the capabilities needed to shift maintenance execution from a schedule-based to a need-based paradigm. Our concept (**Fig. 1**) leverages additive manufacturing to embed small magnetoelastic and thermomagnetic sensing elements that undergo changes in magnetic permeability in response to mechanical or thermal stimuli, modulating the magnetic flux through detection coils placed on or near the surfaces of the printed parts. In addition to providing accurate, in-situ damage assessments of structural health in difficult-to-access locations, our sensing method is wireless, compatible with non-contact readout, battery-free on the embedding side, and it can easily be integrated with manufacturing technologies that are commonly used in industry.

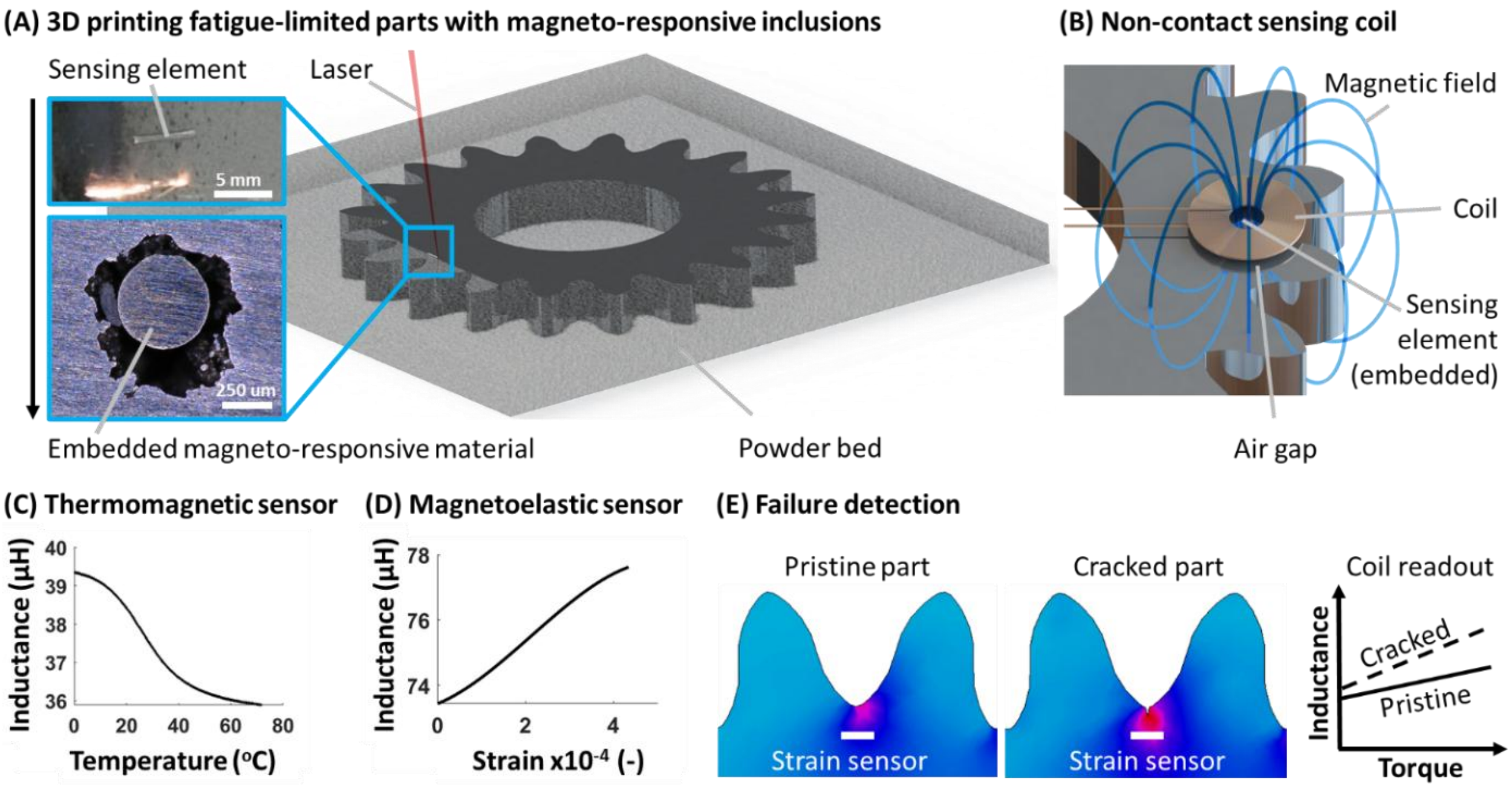

***Fig. 1. Structural health monitoring approach based on wireless, non-contact strain and temperature sensing using embedded magneto-responsive materials.*** ***(A)*** *Sensing elements containing magnetoelastic or thermomagnetic materials are embedded during additive manufacturing of a fatigue-limited part.* ***(B)*** *Strain and temperature are measured wirelessly by placing AC coils near the part's surface to generate a magnetic field and record the coil inductance, which is modulated by changes in the magneto-responsive materials' permeabilities due to mechanical or thermal effects.* ***(C-D)*** *The magnetic permeability of Monel and Galfenol inclusions change in response to variations in temperature and strain, respectively.* ***(E)*** *The*

*initiation of a crack changes the stress field in a gear, modulating the permeability of the magnetoelastic inclusion.*

Several magnetoelastic and thermomagnetic materials exist, offering different measurement ranges and sensitivities *(13–16)*. In this study, we use Galfenol (FeGa) inclusions to measure strain and Monel (NiCu) inclusions to measure temperature. The magnetic permeability of Galfenol changes with stress via reorienting magnetic domains, and the permeability of Monel varies as it approaches its Curie temperature. AC coils placed at the part's surface generate a 1 kHz magnetic field and measure an induced impedance which is modulated by changes in the inclusions' permeabilities.

The layer-by-layer process of additively manufacturing a part can be interrupted to insert a sensing element anywhere within the structure. Embedding sensors below the part surface provides proximity to critical stress locations while protecting the sensor from harsh operating conditions, tear-off, and environmental degradation. To shield the sensing elements from the melt pool in laser powder bed fusion as well as the associated high temperatures and thermal gradients, the magneto-responsive materials can be encapsulated within a material similar to the alloy being printed.

This paper discusses the scientific principles that underpin non-contact sensing of temperature and strain below part surfaces based on magneto-responsive inclusions. We demonstrate embedding magneto-responsive sensing elements while additively manufacturing high strength aluminum parts, measuring strain and temperature with high accuracy and repeatability using standard off-the-shelf electromagnetic inspection hardware. Finally, we leverage strategic sensor placement to detect the onset of plasticity and fatigue-driven crack growth several thousand cycles before failure without accelerating the process.

**Sensing with Magnetoelastic and Thermomagnetic Materials**

Magnetoelastic materials, such as Galfenol (FeGa), Metglas®, cobalt ferrite ($CoFe_2O_4$), and Terfenol-D (TbDyFe) change magnetic permeability under stress due to the reorientation of magnetic domains until their magnetization reaches saturation limits. These saturation limits establish the range of strain measurable by these magnetoelastic elements: this varies from approximately $60\ \mu\epsilon$ for Metglas to approximately $1000\ \mu\epsilon$ for Terfenol-D. Here, we use Galfenol, as it displays a broad sensing range (approximately $400\ \mu\epsilon$) and is easy to machine and manipulate.

Considerable effort has been devoted to studying the magnetoelastic coupling of Galfenol via experiments *(17–20)*, analytical modeling *(20, 21)*, finite elements *(22)*, and density functional theory *(23)*. The material displays high tensile strength (around 500 MPa), high magnetic permeability, and its active properties do not degrade with cycling *(18, 19)*, making it an ideal choice of sensing material for stress monitoring. In the linear regime surrounding a fixed magnetic field and stress state, its constitutive behavior can be described via the piezomagnetic equations (see Supplementary Material) *(19)*. However, monitoring large variations in stress alters Galfenol's permeability in a nonlinear manner (**Fig. 2A**) *(19)*. This nonlinearity requires the adoption of models that capture the stress-induced modulation of Galfenol's permeability *(21)* or the calibration of a model which relates strain to changes in the inductance of a sensing coil directly, which is our approach in this paper.

Crucially, Galfenol's permeability is insensitive to changes in temperature much lower than its Curie temperature (approximately $700\ ^{\circ}\mathrm{C}$), allowing strain measurements to be decoupled from thermal effects. Vibrating sample magnetometer (VSM) measurements demonstrate this property in the 25 $^{o}$C to 100 $^{o}$C temperature range (**Fig. 2B**). However, a coefficient of thermal expansion (CTE) mismatch between the magnetoelastic inclusion and the parent metal in which it is embedded can introduce elastic strains $\varepsilon_{\Delta T,el}$ that bias the measurement. The effect can be decoupled analytically in the linear elastic regime using the relation:

$$\varepsilon_{\Delta T,el} = (\alpha_{FeGa} - \alpha_p)\Delta T \, , \quad (1)$$

where $\alpha_{FeGa}$ and $\alpha_p$ are the linear CTEs of Galfenol and the parent material, respectively, and $\Delta T$ is the change in temperature. This calibration requires measuring the temperature of the magnetoelastic inclusion using a thermomagnetic inclusion placed in its vicinity, allowing for independent readout of temperature and strain.

Thermomagnetic material options include NiFe (Permalloy/Mumetal), NiZn ferrites, NiCu (Monel), AlNiCo and FeCo. The temperature sensing range can be tailored by the choice of material and by tuning alloy compositions *(24)*. We use Monel in this study, as it displays high thermomagnetic sensitivity near room temperature owing to its relatively low Curie temperature (approximately 100 $^{o}$C). VSM measurements shown in **Fig. 2C** show the change in Monel's magnetic permeability in the 25 $^{o}$C to 100 $^{o}$C range and highlight the material's low hysteretic losses through the magnetization and demagnetization loops. This is evident from the material's near-zero coercivity. A transition to the paramagnetic state is observed at 100 $^{o}$C. Monel does not undergo changes in magnetic permeability in response to mechanical strains, allowing temperature measurements to be decoupled from the part's strain field.

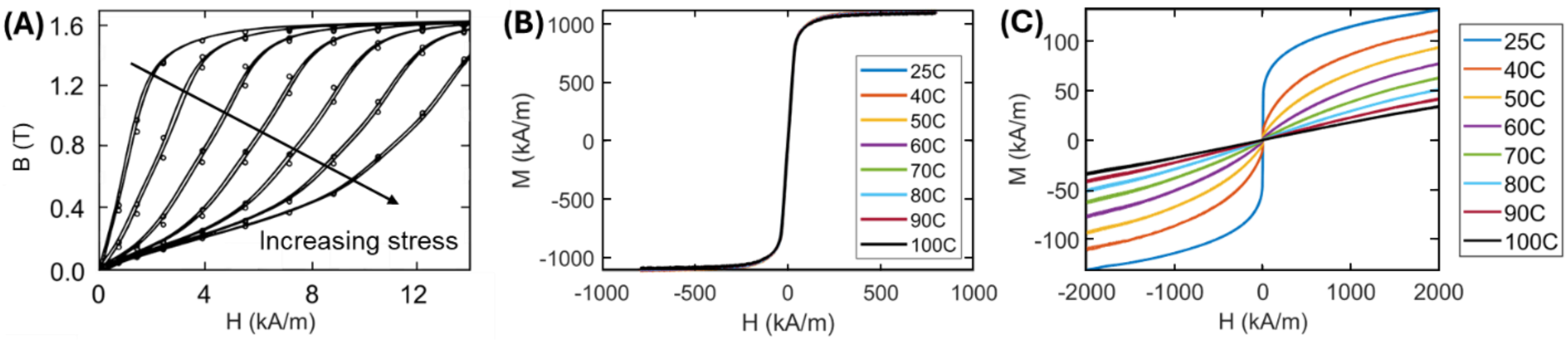

***Fig. 2. Magnetoresponsive properties of Galfenol and Monel. (A)*** *Magnetoelastic behavior of Galfenol, showing changes in the material's permeability in response to stress (19). Reproduced in accordance with NASA Images and Media Usage Guidelines.* ***(B)*** *Vibrating sample magnetometer (VSM) measurements showing the thermal insensitivity of free-standing Galfenol's magnetic properties.* ***(C)*** *VSM measurements highlight the thermal sensitivity of Monel's magnetic properties.*

## Magneto-Responsive Sensor Embedding and Accuracy

Two key differentiators of our structural health monitoring approach relative to other in-situ techniques are that embedding the sensing elements in the interior of a part protects them from tear-off and degradation caused by harsh environment exposure, and the sensing elements do not require wiring to the surface. We use Galfenol sensors with a square 0.5 mm by 0.5 mm cross-section and a length of 5 mm for strain measurements in the direction of their long axis. Monel wire with a 0.5 mm diameter and 5 mm length was selected for temperature measurements. To shield the sensing elements from the extreme thermal environment of Laser Powder Bed Fusion 3D printing of high-strength 7A77 aluminum alloy *(25)*, the magneto-responsive materials were inserted into aluminum microtubes with 1 mm outer diameter and 0.8 mm inner diameter. During 3D printing, not melting the powder in selected areas creates cavities for sensor placement (**Fig. 3A**, **Fig. S1**). When the build resumes, aluminum powder fuses to the aluminum microtube capsules (**Fig. 3B, Fig. S2**). This establishes a good mechanical and thermal connection between the sensor housing and the part. These small magneto-responsive elements do not reduce part lifetime when strategically placed to not intensify stresses in fatigue-critical locations. Crucially for many applications, the sensors continued to function after hot isostatic pressing (HIP) to $400\ ^{o}\mathrm{C}$ and $30{,}000$ psi for two hours.

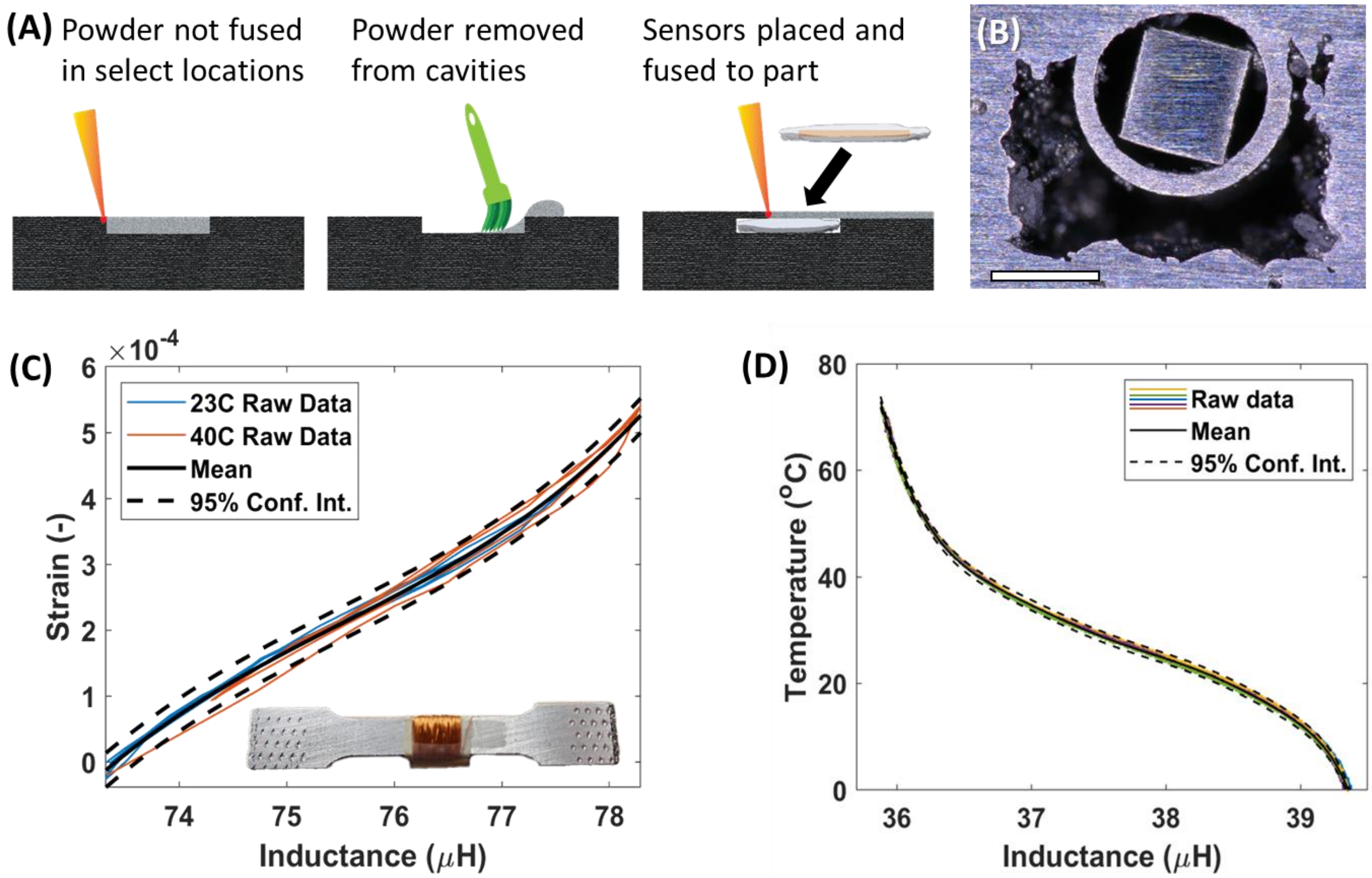

***Figure 3. Embedding magneto-responsive sensing elements. Strain and temperature sensing accuracy.*** ***(A)*** *During 3D printing, several layers of powder are not melted in selected locations, the loose powder is removed to create cavities, and sensing elements are placed. The next layer of powder is spread over the sensing elements and they are welded to the part when the print continues.* ***(B)*** *A cross-section of an embedded Galfenol sensor and its protective capsule. Scale bar: 500 μm.* ***(C)*** *Strain sensing is accurate to* $\pm 27 \times 10^{-6}$ *over a* $6 \times 10^{-4}$ *strain range with copper wire wrapped around a tensile coupon with a galfenol inclusion. Equation 1 yields an analytical CTE mismatch adjustment.* ***(D)*** *Temperature sensing is accurate to* $\pm 0.75\ ^{o}C$ *over a* $70\ ^{o}C$ *range. Both accuracy assessments are relative to a 95% confidence interval.*

The temperature and strain sensing accuracies are dependent on the volume of embedded magneto-responsive material, the embedding depth, coil design and magnetic field strength. Because the focus of this work is not on coil design and we use off-the-shelf components for wireless, non-contact sensing, we provide an accuracy assessment that isolates the contribution of the magneto-responsive materials by wrapping copper wire around printed coupons. **Fig. 3C** shows that strain sensing is accurate to $\pm 27 \times 10^{-6}$ over at least a $6 \times 10^{-4}$ strain range (to a 95% confidence interval). This accuracy represents data collected over four loading and unloading cycles to 70 MPa at both $23\ ^{o}\mathrm{C}$ and $40\ ^{o}\mathrm{C}$, using Equation 1 to decouple the thermal and mechanical effects. This data was collected at thermal equilibrium inside an environmental chamber with a thermocouple adhered to the surface of the specimen. **Fig. 3D** shows that temperature sensing is accurate to $\pm 0.75\ ^{o}\mathrm{C}$ over a $70\ ^{o}\mathrm{C}$ range (to a 95% confidence interval) over five heating and cooling cycles, with the inductance measurement calibrated to a thermocouple embedded in the coupon (**Fig. S3**).

**Wireless, Non-Contact Sensing**

In this first demonstration of wireless, non-contact sensing, we accurately measure strain and temperature by embedding Galfenol and Monel inclusions beneath the surface of a specimen under both tensile and compressive loading. **Fig. 4A-B** shows two coils (TDK Corporation) suspended by a 3D printed mount 0.5 mm from the surface, querying the magnetic permeability of sensing elements embedded 2.5mm beneath the surface. **Fig. 4C** shows the raw data collected over 6 loading cycles at each of three temperatures (23 ºC, 30 ºC, and 40 ºC). CTE mismatch effects are decoupled using Equation 1, yielding the sensing accuracy of $\pm 77 \times 10^{-6}$ over a $1.7 \times 10^{-3}$ strain range. (**Fig. S4** displays the raw data prior to the CTE adjustment.) This high accuracy could be improved even further by using optimal sensing coils.

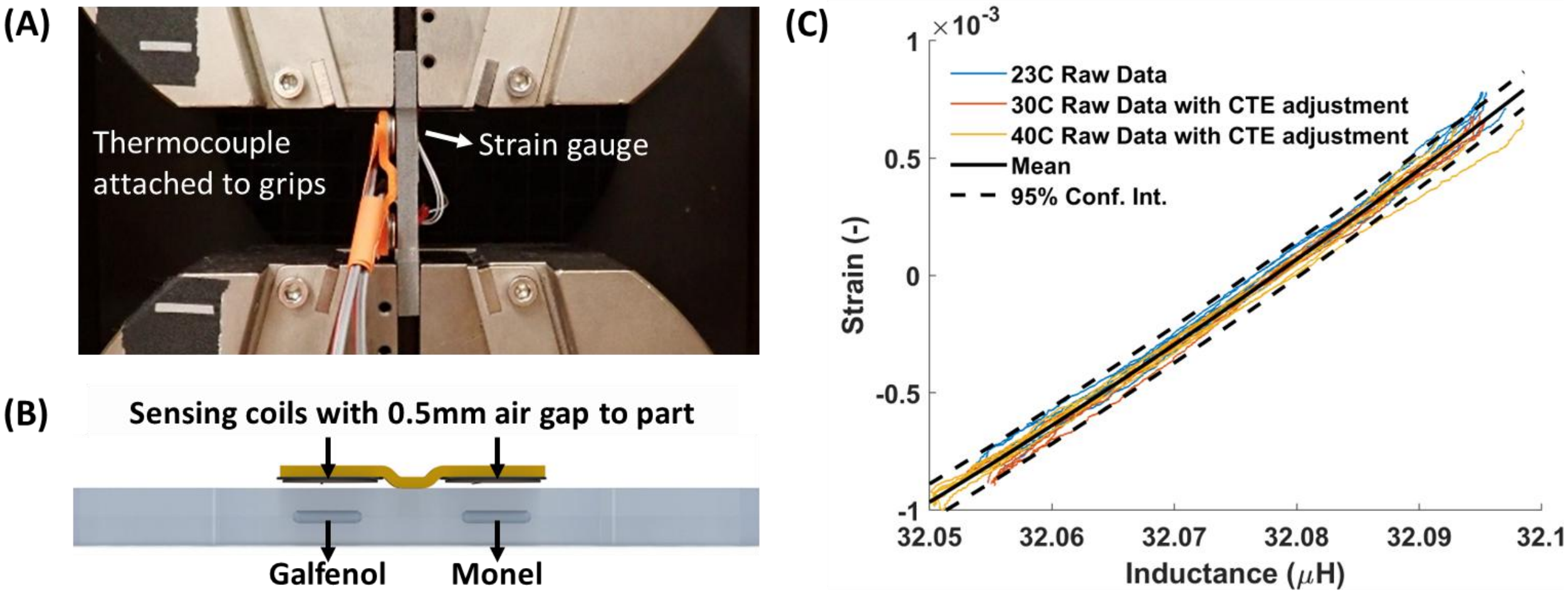

***Figure 4. Non-contact strain measurement accuracy. (A)*** *Experimental setup for non-contact strain measurements which includes two suspended AC sensing coils, as well as a thermocouple and a strain gauge for signal calibration and comparison.* ***(B)*** *Tensile coupon cross-section, illustrating coils suspended 0.5 mm above the surface of the specimen, with magnetoelastic (Galfenol) and thermomagnetic (Monel) inclusions 2.5 mm below the part surface.* ***(C)*** *Non-contact strain measurement showing 6 load cycles at each of three temperatures, using Equation 1 for CTE mismatch adjustments. This configuration's non-contact measurement accuracy is* $\pm 77 \times 10^{-6}$ *over a* $1.7 \times 10^{-3}$ *strain range.*

**Wireless Monitoring of Plasticity and Crack Growth**

One compelling use for this wireless sensing approach is to monitor the onset and development of structural damage. Here, we demonstrate that our wireless sensing method can detect the onset of plasticity as well as the formation and sub-critical propagation of cracks in key locations. These capabilities were demonstrated using 3-point bend coupons with embedded Galfenol and Monel sensing elements (**Fig. 5A**, **Fig. S5**). For the plasticity detection experiment shown in **Fig. 5B**, an 8 mm thick, 24 mm wide, 54 mm long coupon was subjected to small loading and unloading increments. The Galfenol sensing element was embedded 2mm from the outer surface, ensuring that the stress at the inclusion was considerably lower than on the outer fiber of the parent material. The onset of plasticity in the parent material is indicated by the deviation of the loading and unloading paths.

To demonstrate wireless sensing for monitoring fatigue-driven sub-critical propagation of cracks, sensors were placed in 8 mm thick, 12 mm wide, 54 mm long coupons at varying distances from the outer fiber (directly above the point of maximum tensile stress in the beam). The coupons were subjected to force-controlled cycles (10Hz loading from 25MPa to 250MPa at the point of maximum tensile stress). The scanning electron microscope (SEM) image in **Fig. 5C** shows the crack nucleation site at the point of maximum tensile stress on the coupon surface (dashed contour) and fatigue-driven radial crack growth. As cracks initiate and grow, coupons became more compliant, increasing the strain transduced to the embedded sensing elements and modulating the magnetoelastic material's magnetic permeability. Crucially, **Fig. 5D** shows that crack growth is consistently detected thousands of cycles before critical failure.

Care needs to be taken with the location where the sensors are placed to not induce severe structural property debit. Fatigue tests on ten baseline coupons without embedded sensing elements survived $33{,}485 \pm 12{,}537$ cycles under the same loading conditions as the coupons with magneto-responsive inclusions. The specimens of **Fig. 5D** with sensing elements embedded at least 2.5mm from the critical location survived to within one standard deviation of the baseline average durability, indicating that crack growth can be monitored without severely affecting fatigue life so long as the embedded sensors are placed at a sufficient distance from the critical location.

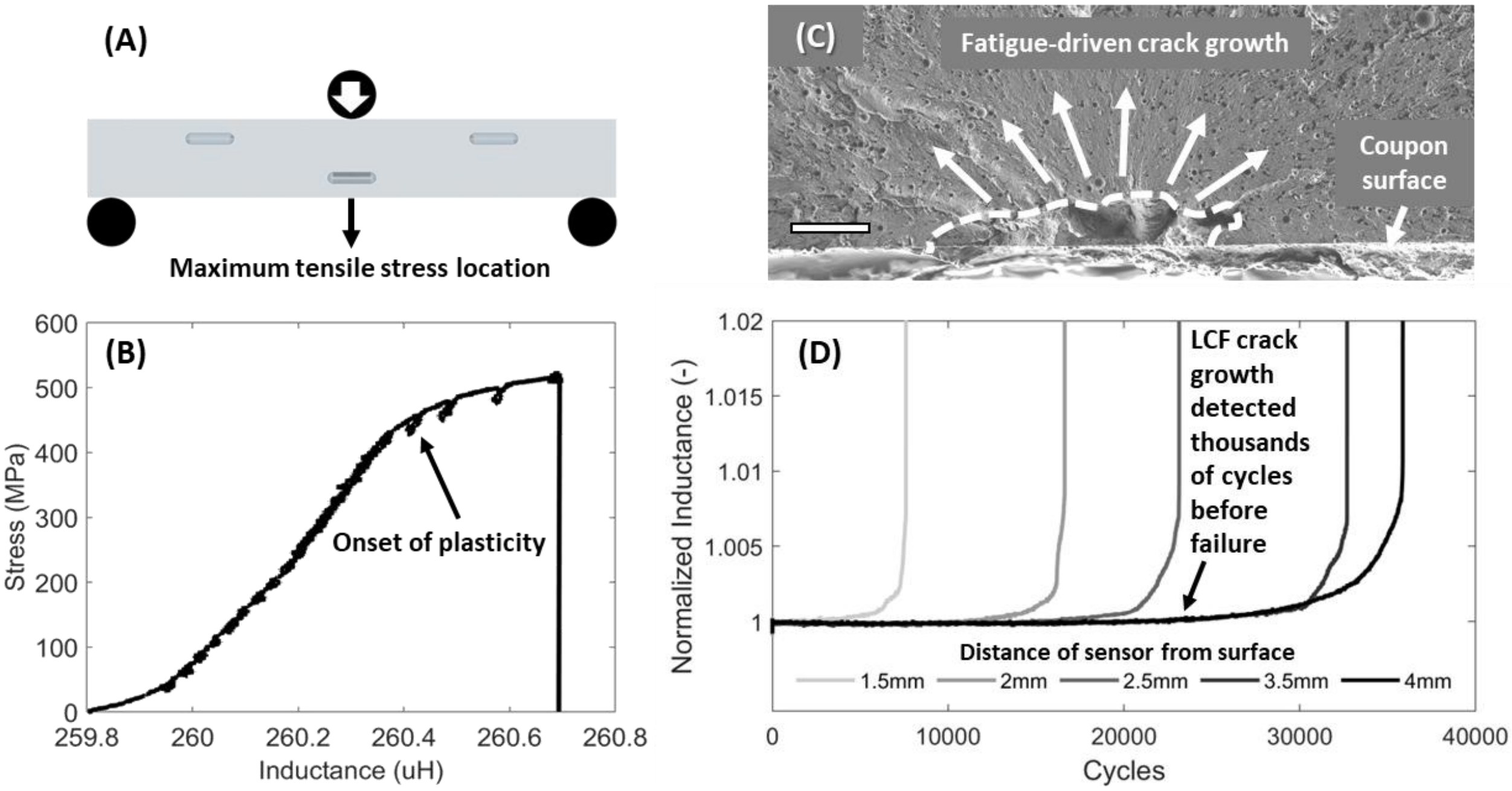

*Figure 5. **Wireless monitoring of plasticity and crack growth. (A)** Cross-sectional illustration of a 3-point bend test specimen with embedded sensing elements (for example, Galfenol in the left and center cavities and Monel in the right cavity). **(B)** Maximum outer fiber stress vs. coil inductance shows the wireless detection of plasticity during a 3-point bend test using a surface-mounted coil. The deviation of the loading and unloading paths in the stress vs. inductance curve indicates the onset of plasticity in the parent material. **(C)** Fatigue test coupon with the crack nucleation site (dashed contour) at the point of maximum tensile stress on the coupon surface*

*and fatigue-driven radial crack growth. Scale bar: 200 µm.* ***(D)*** *Wireless monitoring of low-cycle fatigue-driven crack growth. Coupons become more compliant as cracks grow, transducing more strain to the sensing element, which modulates the magnetic field generated by coils mounted on the surface of the coupons.*

## Conclusions

This approach can be extended to other alloys, including titanium, nickel, and steel, noting the importance of the protective sensor housing for high-melting point alloys. Improvements to the sensing coils would allow miniaturizing the embedded elements, minimizing the structural property debit. The maximum embedding depth attempted in this study was 4mm. This still yielded successful crack growth monitoring (Fig. 5D), but the sensing depth of the system can be increased by lowering the frequency or increasing the applied power of the emitted alternating field, as well as by optimizing the coil design (e.g., by including a high-permeability material such as Permalloy within the coil core). This would also enable the usage of smaller inclusions and improve the sensitivity of the system at shallower embedding depths.

Our approach could overcome current barriers to widespread adoption of structural health monitoring during service, preventing catastrophic failures and shifting the execution of maintenance from a schedule-based to a need-based paradigm. Furthermore, this technology also enables the collection of performance data during service to inform future designs. We envision that this non-contact method of probing stress and strain will be especially useful for monitoring the structural health of rotary components in helicopters, power plants, and a broad set of turbomachinery.

## Acknowledgements

Funding was provided by HRL Laboratories, LLC and the Defense Advanced Research Programs Agency under the Structural Evaluation through Non-contact Sensor Embedding (DARPA SENSE) program (agreement number HR00112490321) managed by Dr. Andrew Detor. The authors gratefully acknowledge informative discussion with Geoff McKnight, Mark O'Masta, Lorenzo Valdevit, Mahsa Amiri, Alireza Kermani, Paul Witherell, and Kyle Rosenkoetter. This research was, in part, funded by the U.S. Government. The views and conclusions contained in this document are those of the authors and should not be interpreted as representing the official policies, either expressed or implied, of the U.S. Government.

## Conflict of Interest Statement

The authors declare that they have no known competing financial interests or personal relationships that could have appeared to influence the work reported in this paper. HRL Laboratories, LLC has filed U.S. Patent Application Number 18/883,761 with the U.S. Patent and Trademark office.

## Data Sharing Statement

The data that support the findings of this study are available from the corresponding author upon reasonable request.

# Supplementary Materials

## Materials and Methods

Aluminum 7A77.50, a novel 7000 series high strength aluminum alloy developed by HRL *(25)* was chosen as the feedstock powder. Crack-free, dense parts were printed in a Renishaw 500Q metal additive manufacturing (AM) system. Aluminum microtubes from Albion Alloys were used to encapsulate the magneto-responsive inclusions. Print parameters and void geometry can be further developed to reduce wall roughness and conform to the microtube exterior more precisely (**Fig. S2)**. Other alloys such as AlSiMg could be used as well.

Galfenol sensors were cut and ground from bar stock. Because mechanical forces must be transduced from the parent material to the strain sensing element, the ends of the microtube capsules are crimped onto the Galfenol using pliers to form a firm mechanical connection between the two components.

Magnetic properties of individual Galfenol and Monel samples were measured using a Microsense EZ9 vibrating sample magnetometer (VSM). Samples were affixed to a sample rod and oriented such that the long axis was parallel with the applied magnetic field. The sample is set to vibrate at a prescribed frequency and the induced voltage in a pair of pick-up coils was detected using a lock-in amplifier. The magnetic field was swept from positive to negative to positive saturation (maximum field = 2.9 T), resulting in a full hysteresis loop. A full hysteresis loop was measured at constant temperature. Samples were allowed to reach equilibrium at the prescribed temperature prior to beginning measurements.

Keysight E4980A Precision LCR Meters were used to generate the electrical currents through the coils (1 V level, 10 mA) and monitor their impedance while thermally or mechanically driven changes in permeability of the magneto-responsive inclusions modulate the induced magnetic fields. Off-the-shelf sensing coils (TDK Corporation, 11 mm outer diameter, 1 mm height, 31.8 μH nominal inductance) generated the magnetic fields for the measurements used in Figs. 4 and 5. To reduce cross-talk in specimens with multiple sensing elements, one coil was driven at 1 kHz, and the other at 1.1 kHz.

After interrupting the print process, the powder was loosened with a brush and blown away before sensing elements were manually placed in the cavities **(Fig. 3A, Fig. S1)**. This sensor placement could be automated in production-scale applications. Next, a new layer of powder was spread and lased around the sensing elements to fuse the aluminum powder to the aluminum microtube capsules. This establishes a good mechanical and thermal connection between the sensor housing and the part **(Fig. 3B, Fig. S2)**.

Fracture surface analysis (Fig. 5C) was done utilizing a Keyence optical microscope as well as a Hitachi SU7000 SEM to determine fracture initiation sites. All fracture surfaces of the 3-point bend fatigue coupons indicated surface initiation on the tensile side of the specimen, with clear radial patterns emanating from the initiation point on the surface of the coupons. In Fig. 5D, the inductance is normalized relative to the initial reading in each coupon's mounted coil to facilitate visualization across samples with sensors embedded at varying distances from the surface.

## Supplementary Text

In the linear regime surrounding a fixed magnetic field and stress state, Galfenol's constitutive behavior can be described via the piezomagnetic equations described below *(19)*:

$$\vec{B} = [\mu^T]\vec{H} + [d^{*H}]\vec{T}$$
$$\vec{S} = [d^T]\vec{H} + [s^H]\vec{T} \quad (S1)$$

where $\vec{B}$ is the magnetic flux density, $\vec{H}$ is the magnetic field, $\vec{S}$ is the strain, $\vec{T}$ is stress. The piezomagnetic coefficient tensors are represented by $[d^{*H}]$ and $[d^T]$, while $[s^H]$ is the compliance tensor and $[\mu^T]$ is the magnetic permeability tensor. The superscripts $H$ and $T$ indicate values measured at constant magnetic fields and stresses, respectively.

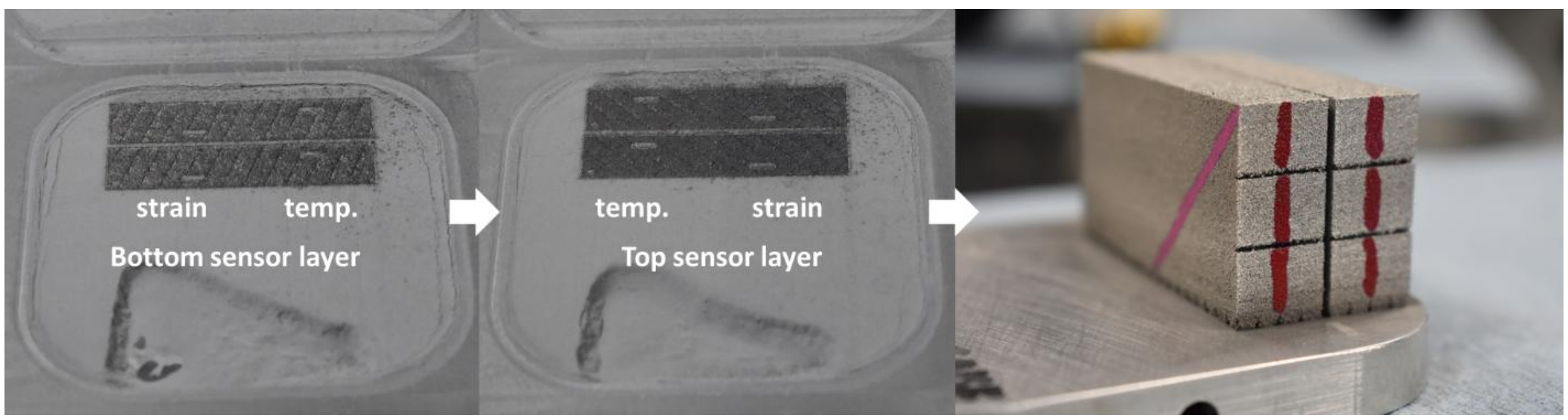

**Fig. S1.**

Intermediate and final stage of Laser Powder Bed Fusion (LPB) of coupons with magnetoelastic and thermomagnetic inclusions. The builds are divided into individual coupons using a wire EDM cutting process.

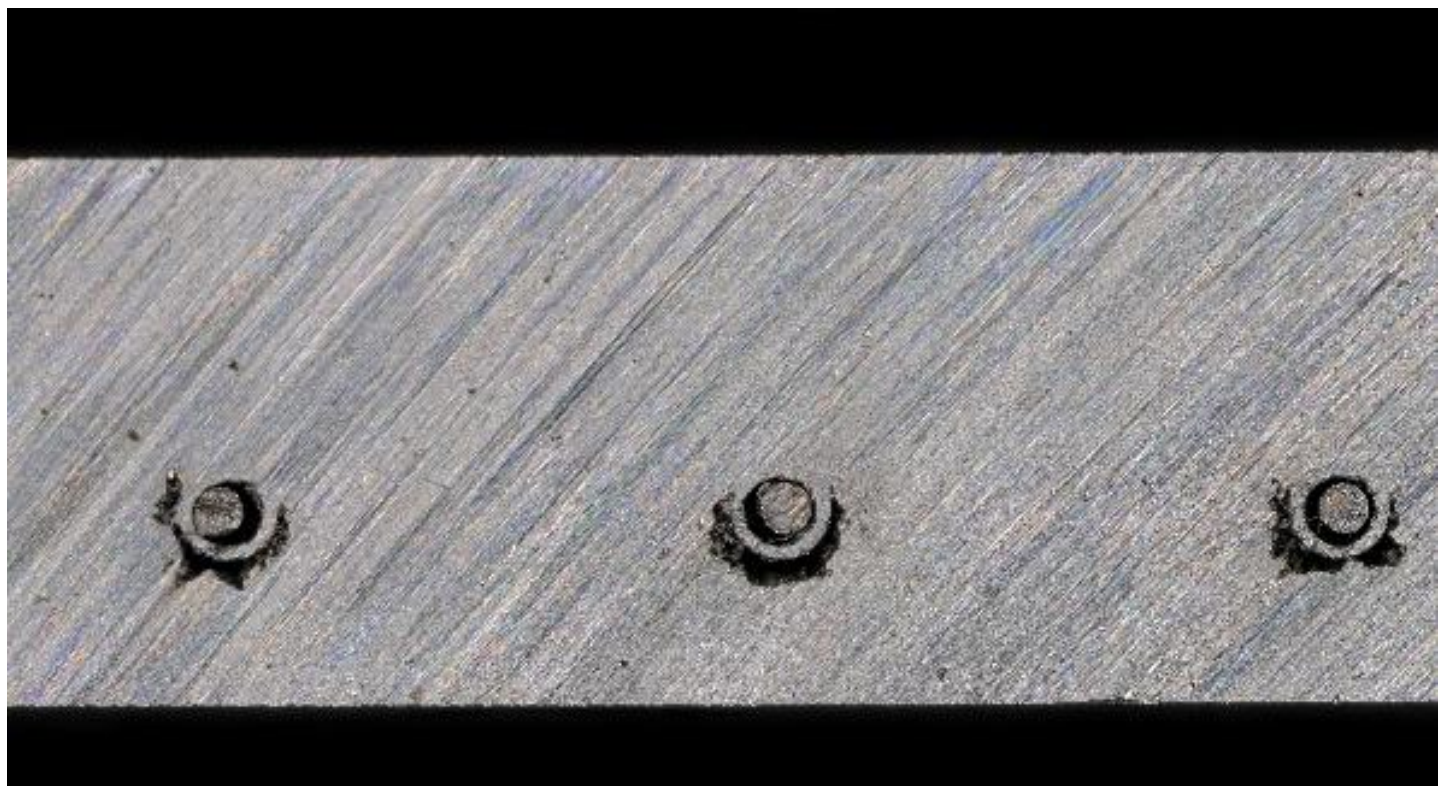

**Fig. S2.**

Cross-section of process development coupon showing examples of cavity geometries swept for tuning the embedding process.

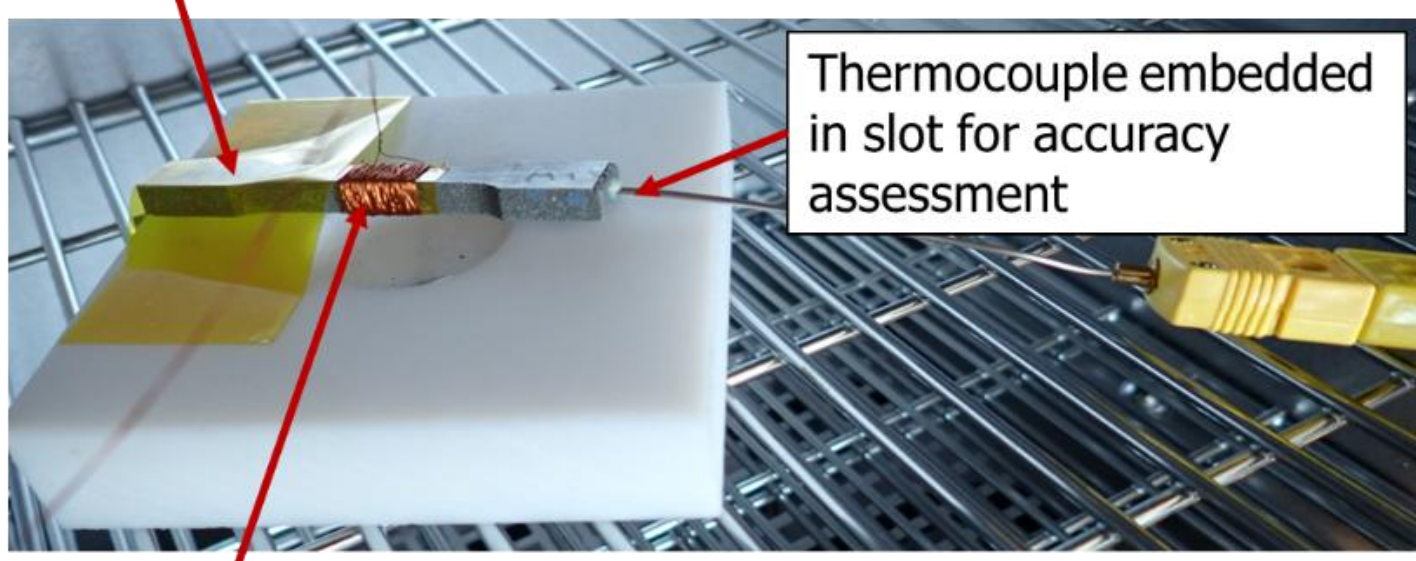

**Fig. S3.**

Experimental setup for assessing temperature measurement accuracy of thermomagnetic monel inclusions inside a thermal chamber. A thermocouple was embedded within the coupon to serve as reference for the modulated coil inductance.

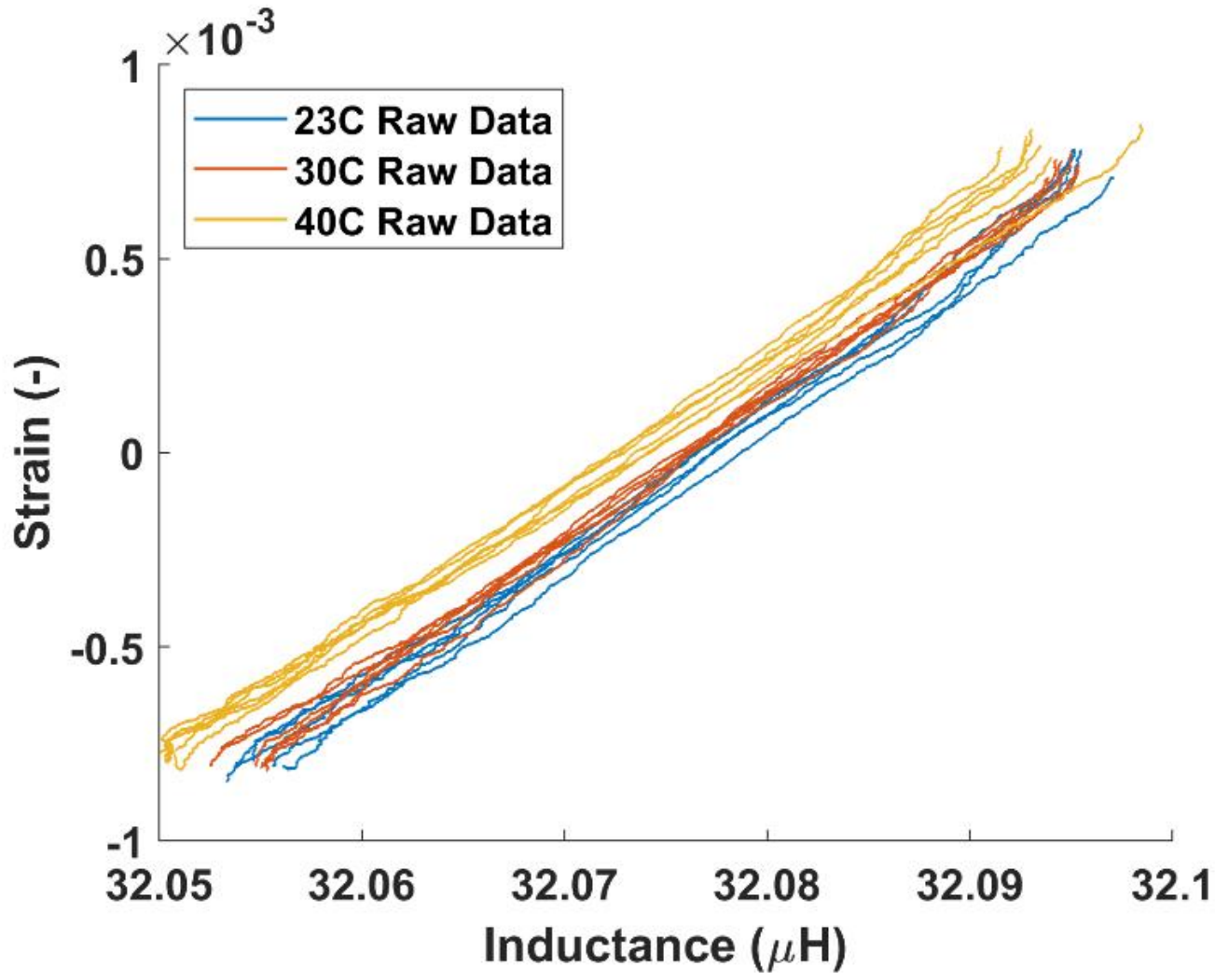

**Fig. S4.**

Raw data collected over 6 loading cycles at each of three temperatures (23 ºC, 30 ºC, and 40 ºC) via wireless, non-contact sensing, prior to the CTE mismatch adjustment performed using Equation 1 to yield the accuracy assessment discussed in Fig. 4C.

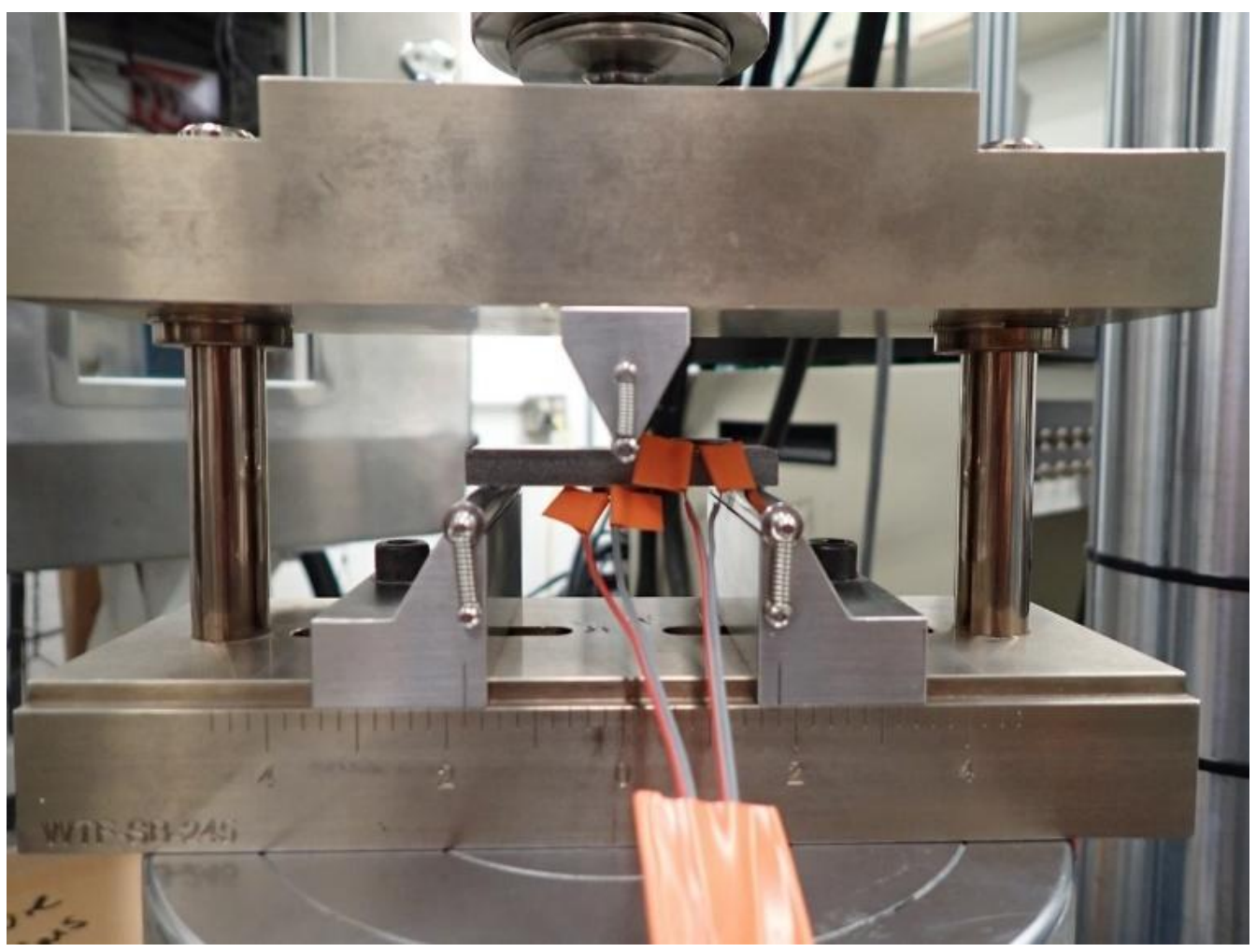

**Fig. S5.**

Experimental setup for conducting 3-point bend low cycle fatigue tests. The coupons display Galfenol coupon at the center of the beam’s axis with varying distances from the coupon’s bottom surface where maximum tensile stresses occur.